\documentclass{he_symp}
\usepackage{psfig,graphicx,epsfig}
\usepackage{color}
\usepackage{amsmath,amssymb,epic,eepic,array}
\begin{document}
\renewcommand{\FirstPageOfPaper }{ 200}\renewcommand{\LastPageOfPaper }{ 205}

\title{Emission altitude in radio pulsars}
\author{Jaros\l aw Kijak}  
\institute{Institute of Astronomy, University of Zielona G\'ora,
Lubuska 2, 65-265 Zielona G\'ora, e-mail: jkijak@ia.uz.zgora.pl}

\maketitle

\begin{abstract}
This paper presents a method of estimation of emission
altitudes using observational data - precise measurements of pulse profile widths
at low intensity level. The analysis of emission altitudes obtained using this method
for a large number of pulsars gives constraints that should be useful for theory of
coherent pulsar emission. It seems that radio emission originates at altitudes of
about few percent of the light cylinder and that they depend on frequency~$\nu$,
pulsar period $P$ and period derivative $\dot P$.
\end{abstract}

\section{Introduction}

It is thirty five years since the first pulsar was discovered, but the mechanism 
of coherent radio emission in pulsars is still not well known. 
In the polar cap model (e.g. Goldreich \& Julian 1969, hereafter GJ; 
Ruderman \& Sutherland 1975), the observed emission of radio pulsars is a coherent
radiation of relativistic particles flowing along dipolar field lines.
This radiation is generated not far from the surface of the neutron star (NS).
The emission altitude $r_{\rm em}$, which is the distance from NS
to a place where radiation is generated, is a crucial
parameter of emission region. It is commonly believed  that $r_{\rm em}$ 
dependes on the observed frequency. This view, 
supported by the systematic narrowing of average pulse width with increasing frequency
(Thorsset 1991), is described by a
radius-to-frequency mapping (RFM). It is obvious that all observational limits for emission 
altitude can be crucial for understading
the physical mechanism generating pulsar radiation. 
One of the ways to estimate the emission altitude (Cordes 1978, 1993) involves an analysis of
observational data concerning the average pulse profiles.  This method was adopted and
developed by Kijak \& Gil (1997, 1998, hereafter Papers I and II) and  
Kijak (2001, hereafter Paper III),
using precise measurements of pulse profile width for the largest available data set 
(about 40 PSRs). Three straightforward assumptions were used: 
(1) the pulsar radiation is narrow-band,
with a radius-to-frequency mapping operating in the emission region
(that is, a narrow-band frequencies $\Delta\nu\ll\Delta\nu_{\rm tot}$ 
are emitted at given altitude $r_{\rm em}$, where $\Delta\nu_{\rm tot}\sim 30$GHz 
is a total band of pulsar radiation), 
(2)~the elementary pulsar emission is relativistically beamed 
tangentially to dipolar magnetic
field lines (beaming angle $\theta\sim 1/\gamma$; $\gamma\sim 100$), 
(3)~the extreme profile wings originate at or near the last open 
dipolar field lines.
The estimation of emission altitude is correct (i.e. valid for describing the actual physical 
phenomenon) as long as the above assumptions are true. 
In other words, if we question one of these assumptions, our 
estimation of emission altitude can be questioned too. 
The first two assumptions are in agreement with basic pulsar 
electrodynamics (GJ; Sturrock 1971; Ruderman \& Sutherland 1975)
and with  the observed narrowing of profiles with increasing frequency 
(Thorsset 1991; Mitra \& Rankin 2002). These two
assumptions are commonly accepted. 
In the third assumption, the edge of pulsar radio beam is determined by the
boundary of the GJ polar cap. 
This assumption is not widely accepted and will be discussed details in~\S3.

\begin{figure*}
\centerline{\psfig{file=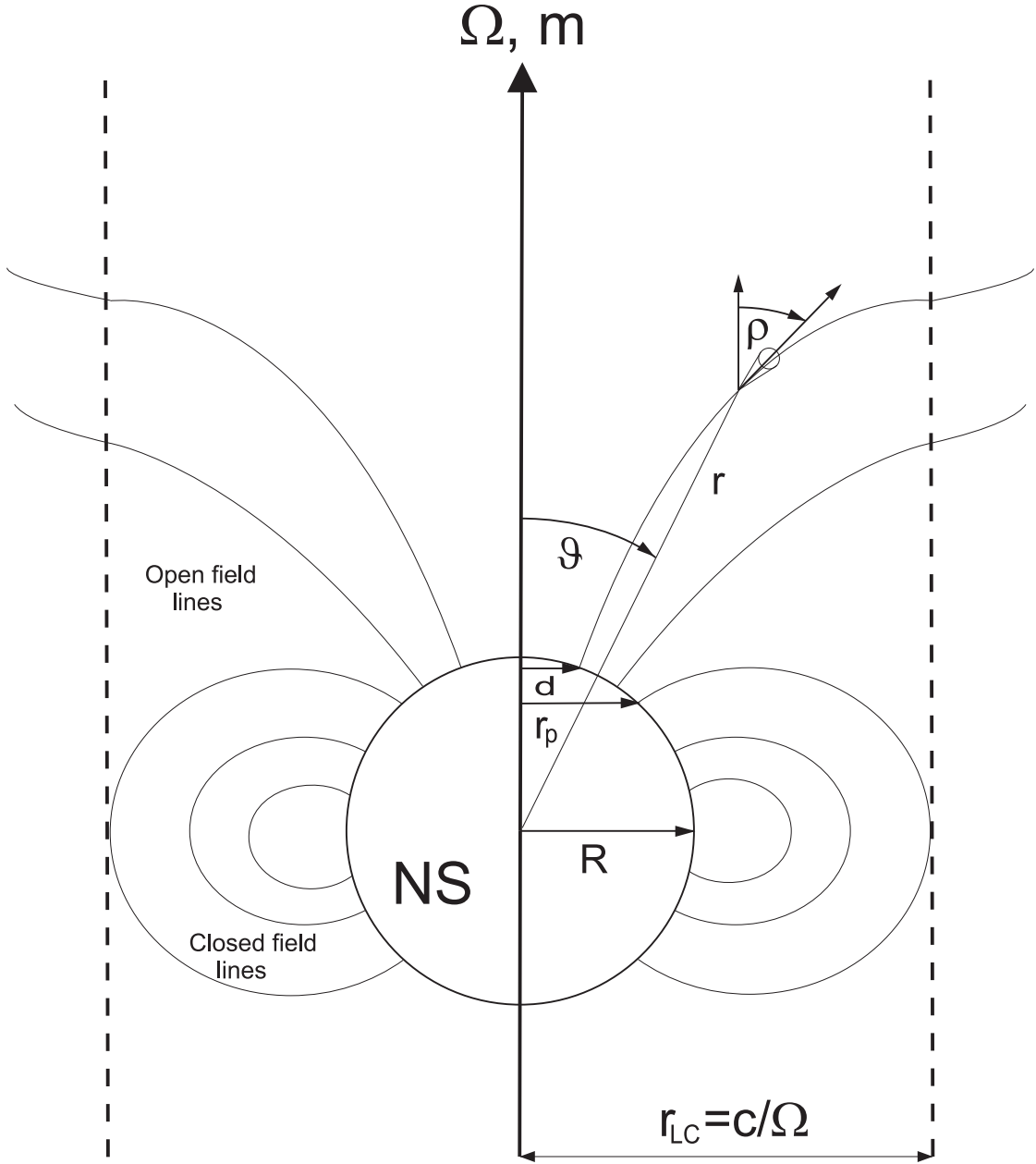,bbllx=1cm,bblly=6.5cm,bburx=20cm,bbury=23.5cm}}
\caption{Geometry of dipole field lines of the neutron star
\label{fig1}}
\end{figure*}

\section{Geometry of emission region}
The pulsar beam originates within the open dipolar field lines at altitudes
much smaller than the radius of the light-cylinder
$r_{\rm LC}=c/\Omega$ (Lyne \& Smith 1998). 
The last open field lines define the polar cap on the neutron star surface. The 
radiating plasma streams outwards along the open field lines.
Let us consider the case of a neutron star with aligned rotation
and magnetic axes (see Fig.~1). 
The equation of dipolar magnetic field lines in polar coordinates
(${\bf r}, \vartheta$) has a form
$r=(R^3/d^2)\cdot\sin{\vartheta}^2$.
For the last open field line 
$\vartheta=\pi/2$ and $r=c/\Omega$, therefore $d=\sqrt{R^3(2\pi/cP)}$ and the radius of
polar cap is 
$r_{\rm p}=1.45\cdot 10^4 P^{-1/2}$[cm], for the neutron star radius $R=10^6$ cm.

\subsection{Geometry on the celestial sphere}
The geometry of pulsar radio emission as defined by Manchester \& Taylor (1977) is presented
in Fig. 2. The rotation and magnetic axes are inclined by an angle $\alpha$. The observer's
angle $\xi=\alpha + \beta$, where $\beta$ is the impact angle of the closest approach of 
the line-of-sight (l-of-s) to the magnetic axis. The beam radius (corresponding to 
a pulse profile) is $\rho\ge\beta$ and the observed
pulse width is $W=2\varphi$, where $\varphi$ is the longitude of the beam edge. 
The fiducial plane containing the rotation $\bf \Omega$ and magnetic $\bf m$
axes defines the so called fiducial phase,
which is usually taken as the longitude $\varphi=0^{\circ}$ (Fig.~2).

\begin{figure}
\psfig{file=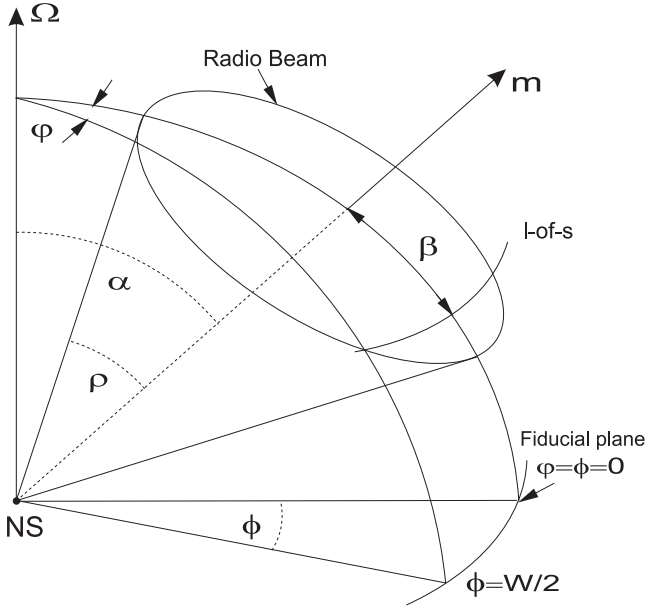,bbllx=6.1cm,bblly=10.5cm,bburx=20cm,bbury=19.5cm}
\caption{Geometry on the celestial sphere. 
\label{fig2}}
\end{figure}

\subsection{The method of estimating emission altitude}
The opening angle of the pulsar beam corresponding to the pulse width $W$ is given by
\begin{equation}
\rho = 2 \sin^{-1}\left(\sin^{2}\frac{W}{4}\sin\alpha \cdot 
\sin(\alpha+\beta) + \sin^{2}\frac{\beta}{2}\right)^{1/2},
\end{equation}
where $\alpha$ is the inclination angle 
between the rotation and magnetic axes and 
$\beta$ is the impact angle (Gil, Gronkowski \& Rudnicki 1984).
This formula is independent of the shape of the pulsar beam (circular, elliptical, 
patchy, etc.) when the beam is symmetric 
with respect to the fiducial plane. This of course assumes that the edge of beam
is close to circular or elliptical in shape.

The opening angle $\rho$ 
expressed above is also the angle between {\bf m} axis
and the tangent to magnetic field lines at points where 
the emission corresponding to the apparent pulse width $W$ originates. For dipolar 
field lines (see Fig.~1)
\begin{equation}
\rho=1.24^{\circ} s (r/R)^{1/2}P^{-1/2}~,
\end{equation}
where $P$ is the pulsar period, 
$r$ is the emission altitude (at which radiation at a given frequency $\nu$ is generated), 
$R=10^6$cm is the neutron star radius and the mapping parameter 
$s=d/r_p$ describes the locus of corresponding field lines on the polar cap
($s=0$ at the pole and  $s=1$ at the edge of the polar cap.
Finally, we calculate the emission altitude as follows
\begin{equation}
r_{\rm em}=R\left(\frac{\rho}{1.24^{\circ}s}\right)^2P~~~.
\end{equation}

\section{Precise measurements of pulse profile and the edge of the radiation beam}
The most important question is what we mean 
by the edge of pulsar beam and the edge of an average pulse profile. 
First of all, the edge of pulsar beam  is related
to the edge of a pulse profile
in the following term: when the pulsar beam
starts passing through the radiotelescope, then
we start observing a weak coherent signal in the receiver. 
The moment at which the pulsar beam enters the radiotelescope corresponds to 
the moment at which the recorded noise rapidly  becames a coherent signal
(and signal becames a noise when the beam leaves the radiotelescope).
In studies on emission regions (Papers I, II and III) we used
precise measurements of the longitude corresponding to 
these events. The egde of the beam  is defined 
as the longitude at which recorded signals are distinguished from the noise
and have the instensity at least 
0.1\% of the maximum intensity. Since the pulsar radiation is relativistically 
beamed along dipolar field lines, 
we can attempt to calculate which field lines should
be tagged as coming from the emission region. 
This should be easiest for the last open magnetic field lines,
which are believed to be associated with the lowest detectable level of radio 
emission - i.e. at the profile wings.

The radio emission from pulsars is characterized by a short time of duration,
typically 3-10\% of its period. 
However, a few objects demonstrate the emission through the entire pulsar period.
Hankins et al. (1993) showed, using the VLA, that the emission through the entire 
pulsar period is observed only  in cases when the inclination angle
$\alpha$ is smaller than the beam radius $\rho$ (see Fig. 2), that is,
the l-of-s is always inside the emission beam (Gil 1985).
In general, there is a transition from the noise to
the weak emission at the extreme profile wings (Fig. 3 and 4). 
In our analysis of emission altitudes (Paper I, II and III),
we measured pulse widths at the extremely low intensity 
level corresponding to about 0.05 per cent of the maximum
intensity (see Fig.~3), using the polar-log-scale technique (Hankins \& Fowler 1986).

It is not certain that the entire GJ polar cap is involved
in the emission process thus, 
the edge of pulsar beam is emitted at or near the last open field lines
(which correspond to $s=1$). It is possible that a forbidden area exists at the outer
parts of the GJ polar cap.
Evidently, we cannot give a theoretical answer to these questions. 
Recently, Gangadhara \& Gupta (2001) found new outermost weak components in 
the pulse window of pulsar B0329+54 at frequencies of 320 and 610 MHz. 
The separation $\Delta\phi$ between the newly detected components is 34$^{\circ}$.
In Fig.~5  the profile of PSR B0329+54 at 320 MHz is presented,
which was used by Kijak \& Gil (1998)
in their measurement of pulse width
and corresponding estimation of emission altitude (Paper II, Fig.~2). 
Let us note that the pulse width measurement $W$ at 0.05\% level is about 49$^{\circ}$.
A comparison of these two measurements ($\Delta\phi=34^{\circ}$ and 
$W=49^{\circ}$) clearly shows that the profile width presented in
Fig.~5 includes new components found by Gangadhara \& Gupta (2001). 
Thus, the reported new outermost components in the profile of PSR B0329+54 do
not contradict our assumption that $s\approx 1$ at the pulse edges.
Our method of profile width measurements has proven to be sensitive to
very weak radio emission at profile wings  originating at the edge
of the beam, most probably close to the last open field lines.
These outermost components  are clearly visible in Fig.~5  
at the level of about 0.1\% of the maximum intensity.

\begin{figure}
\centerline{\psfig{file=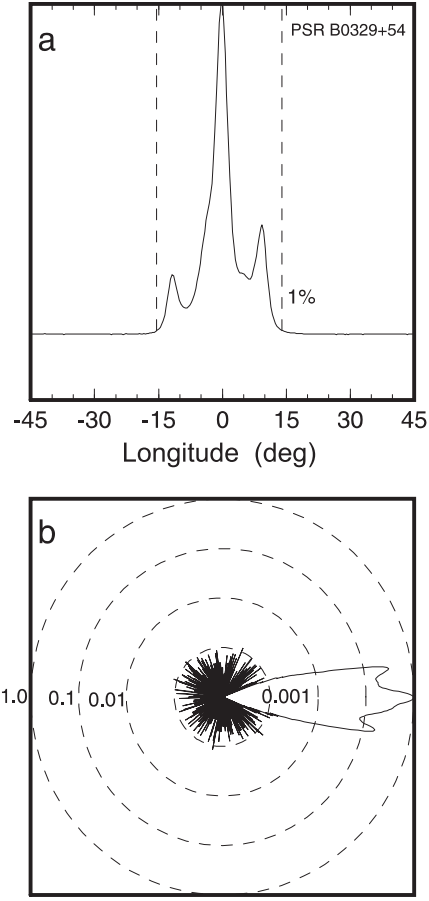,width=8.8cm,clip=} }
\caption{(a) A plot of profile of PSR B0329+54 at 1.4 GHz, (b) the profile of 
the same pulsar in the log-scale and the polar coordinates. The dynamic range is
$10^4$.
\label{fig3}}
\end{figure}

\begin{figure}
\centerline{\psfig{file=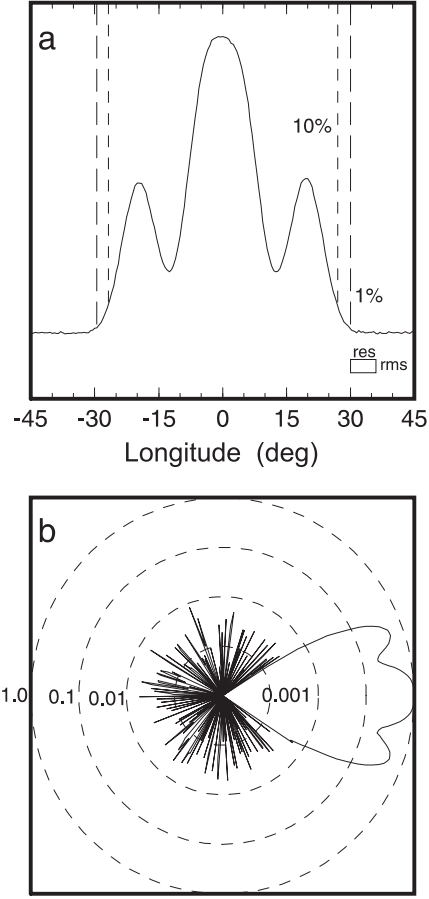,width=8.8cm,clip=} }
\caption{The syntetic data with pulsar signal which comes from open field lines.
The noise was obtained  using a random generator. Error box is marked as
an example (see details in Paper~I).
\label{fig4}}
\end{figure}

\section{Emission altitude estimation}
In our research of radio emission regions we used the high quality
average pulse profiles in the wide frequency range.
We selected pulsars for which some information about 
$\alpha$ and $\beta$ angles was available. 
For this purpose, observations
were performed between 1995 and 1998 using the Effelsberg 100-m radio telescope 
(Lorimer et al. 1998) and the Jodrell Bank data (Gould \& Lyne 1998) were used as well.
As a result, the high quality average profiles were obtained with the signal-to-noise
ratio S/N$ \sim 1000$ in the frequency range between 0.3 and 20 GHz. These profiles
were used to obtain precise measurements of pulse profile widths and to estimate  
the emission altitudes.

\subsection{Results}
The result of emission altitude analysis (Papers I, II and III)  
indicated clearly 
that the emission altitude $r_{\rm em}$ depends not only on the observing frequency
$\nu$, but also on the basic pulsar parameters: pulsar period $P$ and 
period derivative $\dot P$.  
Here, we present the analysis of emission altitudes at 1.4 GHz 
for the largest available data set (37 PSRs). An apparent period dependence
of emission altitudes is visible in Fig. 6, where a formal fit to all data points
gives $r_{\rm em}=(45\pm 3)RP^{0.37\pm 0.05}$. This result confirms the previous analysis of
the emission altitude at different frequencies for a smaller data set 
presented in Papers I and II.
Figure~6 also shows a weak dependence $r_{\rm em}$ on pulsar characteristic age $\tau_6$
(in $10^6$ yrs), which is discussed in Paper III (see its Fig. 2 and Table 2).  
For this analysis, a formal fit to the same data set (37 PSRs) at 1.4 GHz
yields $r_{\rm em}=(33\pm 1)R\tau_6^{-0.07\pm 0.03}$.  In Paper II,
the RFM was derived for 16 pulsars covering the frequency range
between 0.3 and 20 GHz. The average frequency dependence on emission altitude is
represented by the formal fit
$r_{\rm em}\propto \nu^{-0.26\pm 0.09}$. Generally, the pulsar
emission altitude depends on frequency $\nu$, pulsar priod $P$ and pulsar age $\tau$.

\begin{figure}
{\psfig{file=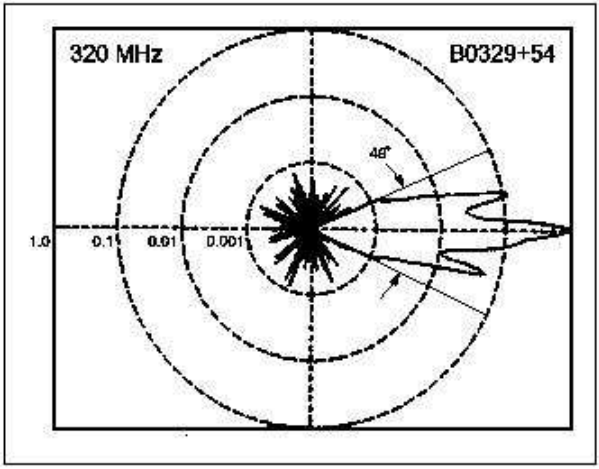,width=8cm
}}
\caption{A pulsar profile of B0329+54 at 320 MHz in the log-scale and 
the polar coordinates. The data come from Jodrell Bank Observatory. 
\label{fig5}}
\end{figure}

\subsection{The semi-empirical formula for emission altitude}
The radial location of the emission regions was discussed in Papers I, II and III.
A general form of the formula for the emission altitude  is
\begin{equation}
r_{\rm KG}={\cal A}R\nu_{\rm GHz}^{-a}\tau_6^{-b}P^{c}~~~,
\end{equation}
where ${\cal A}\sim 55$, $R\cong 10^6$cm, 
$\nu_{\rm GHz}$ is observing frequency (in~GHz), 
$\tau_6$ 
is characteristic age (in $10^6$ yrs), $P$ is pulsar period (in sec). 

The study on emission region in Paper I carried out for
about 10 PSRs at two frequencies $\nu=0.4$ and 1.4 GHz yielded 
${\cal A}=55\pm 5$, $a=0.21\pm 0.07$, $b=0.07\pm 0.03$ and $c=0.33\pm 0.05$. 
Here we verify these values for
a larger data set, as well as for a wider frequency range. 
Taking into account results of \S~4.1, we obtain
\begin{equation}
r_{\rm KG}=(50\pm 10)R \nu_{\rm GHz}^{-0.26\pm 0.09}\tau_6^{-0.07\pm 0.03}
P^{0.37\pm 0.05}~,
\end{equation}
or 
\begin{equation}
r_{\rm KG}\approx(40\pm 8)R \nu_{\rm GHz}^{-0.26\pm 0.09}\dot P_{-15}^{0.07\pm 0.03}
P^{0.30\pm 0.05}~,
\end{equation}
where $\tau_6=16P/\dot P_{-15}$ and $\dot P_{-15}=\dot P/10^{15}$ is the period derivative.
The uncertainties given for various quantities in the above equations follow from systematic
errors in the pulse width $W$ measurements and random errors in estimations of 
the inclination angle $\alpha$ and the impact angle~$\beta$.

\begin{figure}
\centerline{\psfig{file=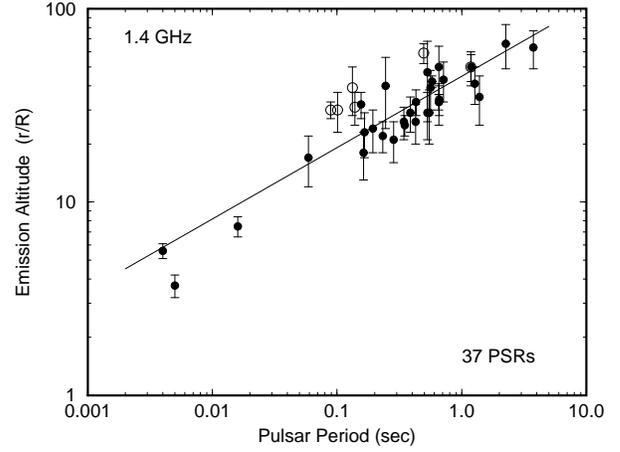,width=8.8cm,clip=} }
\caption{Emission altitude at 1.4 GHz versus pulsar period. Open circles correspond
to young objects (see Paper~III for details). 
\label{fig6}}
\end{figure}

\section{Discussion and Conclusions}
As it follows from Eqs. (5) or (6), 
the radio emission of pulsars originates at the narrow altitude range below
10 per cent of the light-cylinder radius.
In millisecond pulsars, the radio emission region
is much more compact, consistent with their small light-cylinder radii, and is located
closer to the neutron star surface (see Paper II).
Previously, it was thought that millisecond pulsars  have different characteristics of radio
emission from those of normal pulsars. However, it was recently argued (Gil \& Sendyk
2000; Kramer 1999, and this proceedings) that millisecond pulsar emission properties do
not differ from those of typical pulsars.

Recently, Gil et al. (2002) analysed  simultaneous dual-frequency single pulse
observations of PSR B0329+54 at 240 and 610 MHz. 
The phase shifts of the leading and the trailing
conal components are not equal at these frequencies. 
This is caused by the retardation-aberration effects.
In fact, since the conal beams at different frequenies are emitted at different
altitudes, lower frequencies arrive earlier. The retardation-aberration shift of
0.39$^{\circ}$ translates into a difference of emission altitudes of $\Delta r\sim
2.3\times 10^7$cm. This is consistent with the result obtained using the semi-empirical
formula $\Delta r_{\rm KG}=2\times 10^7$cm (Eqs. 5 or 6).

\begin{figure*}
\psfig{file=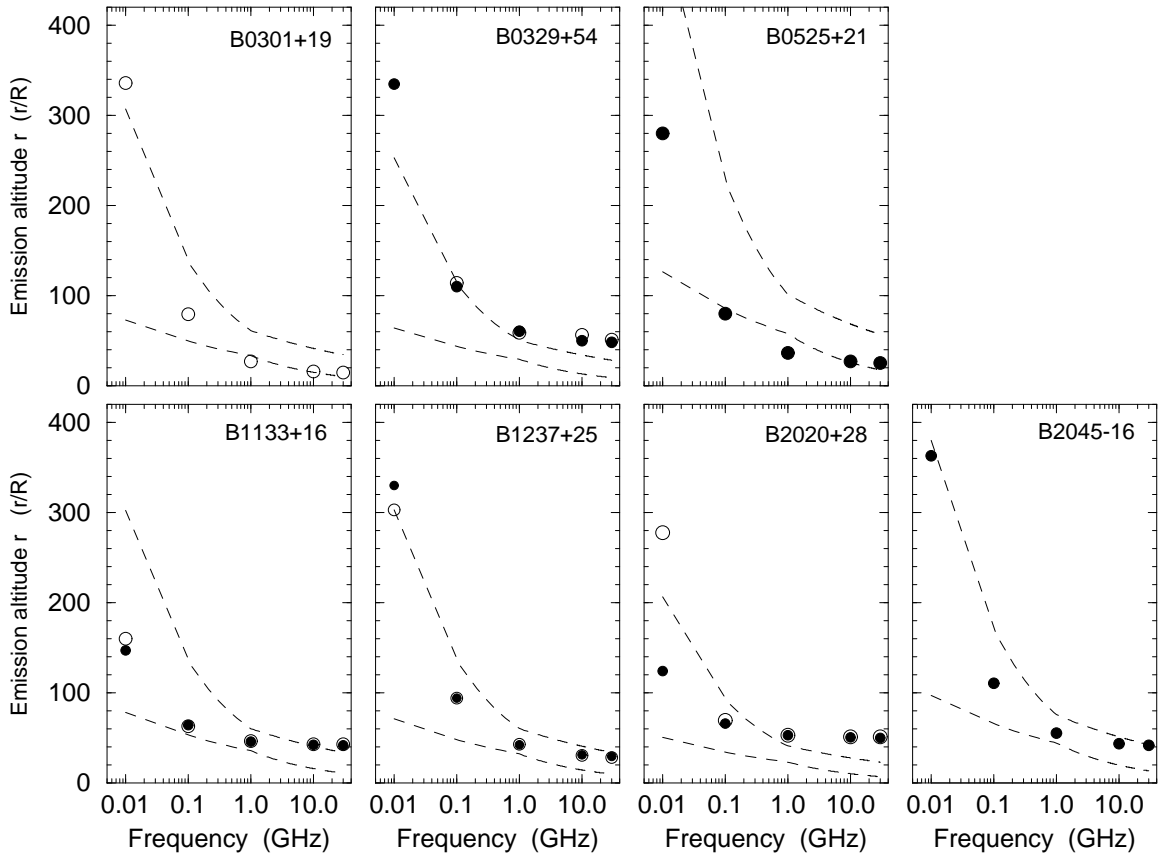,bbllx=2.5cm,bblly=4.5cm,bburx=15cm,bbury=16.8cm} 
\caption{The multifrequency study of emission altitude. Data points represent calculations
from the paper of Mitra \& Rankin (2002). 
The dashed lines show upper and lower limits, calculated
from equation (5), corresponding to limiting values of parameters in the semi-empirical
relationship.
\label{fig7}}
\end{figure*}

Mitra \& Rankin (2002) found the RFM using outer-conal components data. In Fig.~7
we present a comparison between their results and calculation from 
Eqs. (5) or (6). Emission altitudes represented by open and filled circles
were calculated from $r_{\rm em}=h\cdot s^{-2}$, 
where $h$ was the emission height for conal component
and $s$ was taken as the average value from Mitra \& Rankin's paper
for each pulsar, respectively.
The open circles represent calculations for the first group of their
fitted procedure, and filled
circles were calculated using the index of $-2/3$ (see Mitra \& Rankin 2002, their 
Table 4).
Figure~7 shows that the semi-empirical formula describes quite
fr

well the radio emission region in pulsars, consistenly with other results.

The main conclusions from the work on radio emission regions are as follows:
\begin{description}
\item (i) A radius-to-frequency mapping operates in the emission region.
\item (ii) Pulsar radio emission is typically generated at altitudes lower 
than a few per cent
of the light-cylinder radius $r_{\rm LC}$ and the ratio $r_{\rm KG}/r_{\rm LC}$ decreases
with incresing period. The total size of the emission region is below $500R$ in 
longer period pulsars and correspondingly smaller in shorter period pulsars.
\item (iii) The emision region in old pulsars is located at correspondingly
lower altitudes than in young pulsars with approximately the same period.

Qq
  cx\end{description}

\vskip 0.4cm

\begin{acknowledgements}
I gratefully acknowledge the support by the Heraeus foundation.
\end{acknowledgements}


\clearpage


\begin{thebibliography}{} 
\bibitem{c78} Cordes J. M., 1978, ApJ, 222, 1006
\bibitem{c93} Cordes J. M., 1993, in van Riper K. A., Epstein R., Ho C. eds, Proc.
Los Alamos Workshop Cambridge Unov. Press, Cambridge, p. 182
\bibitem{gg01} Gangadhara R. T., Gupta Y., 2001, ApJ, 555, 31
\bibitem{ggr84} Gil J., Gronkowski P., Rudnicki W., 1984, A\&A 132, 312
\bibitem{g85} Gil J., 1985, ApJ 299, 154
\bibitem{gs00} Gil J., Sendyk M., 2000, ApJ, 541, 351
\bibitem{gggk02} Gil J., Gupta Y., Gothoskar P.I., \& Kijak J., 2002, ApJ, 565, 500
\bibitem{gj69} Goldreich P., Julian W. H., 1969, ApJ, 157, 869
\bibitem{gl98} Gould M., Lyne A. G., 1998, MNRAS, 301, 235
\bibitem{hf86} Hankins T. H., Fowler L. A., 1986, ApJ, 304, 256
\bibitem{hmnp93} Hankins T. H., Moffett D. A., Novikow A., Popov M., 1993, ApJ, 417, 735
\bibitem{kg97} Kijak J., Gil J., 1997, MNRAS, 288, 631 (Paper I)
\bibitem{kg98} Kijak J., Gil J., 1998, MNRAS, 299, 855 (Paper II)
\bibitem{k01} Kijak J., 2001, MNRAS, 323, 537 (Paper III)
\bibitem{kll+99} Kramer M., Lange C., Lorimer D. R., Backer D. C., Xilouris K. M., 
Jessner A., Wielebinski R., 1999, ApJ, 526, 957
\bibitem{ljs+98} Lorimer D. R., Jessner A., Seiradakis J. H., Lyne A. G., D'amico N., 
Athanasopoulos A., Xilouris K. M., Kramer M., Wielebinski R., 1998, A\&AS, 128, 541
\bibitem{ls98} Lyne A. G., Graham-Smith F., 1998, Pulsar Astronomy, Cambridge University 
Press, Cambridge, p. 173
\bibitem{mt77} Menchester R. N., Taylor J. H., 1977, Pulsars, Freeman, San Francisco 
\bibitem{mr02} Mitra D., Rankin J.M., A\&A 2002, in press
\bibitem{rs75} Ruderman M. A., Sutherland P. G., 1975, ApJ, 196, 51
\bibitem{s71} Sturrock P. A., 1971, ApJ, 164, 529
\bibitem{t91} Thorsett S. E., 1991, ApJ, 377, 263
\end{thebibliography}
\end{document}